# DNN+NeuroSim V2.0: An End-to-End Benchmarking Framework for Compute-in-Memory Accelerators for On-chip Training

Xiaochen Peng, Shanshi Huang, Hongwu Jiang, Anni Lu and Shimeng Yu
School of Electrical and Computer Engineering, Georgia Institute of Technology, Atlanta, GA
Email: shimeng.yu@ece.gatech.edu

*Abstract*—DNN+NeuroSim is an integrated framework to benchmark compute-in-memory (CIM) accelerators for deep neural networks, with hierarchical design options from device-level, to circuit-level and up to algorithm-level. A python wrapper is developed to interface NeuroSim with a popular machine learning platform: Pytorch, to support flexible network structures. The framework provides automatic algorithm-to-hardware mapping, and evaluates chip-level area, energy efficiency and throughput for training or inference, as well as training/inference accuracy with hardware constraints. Our prior work (open-source framework DNN+NeuroSim V1.1: https://github.com/neurosim/DNN_NeuroSim_V1.1) was developed to estimate the impact of reliability in synaptic devices, and analog-to-digital converter (ADC) quantization loss on the accuracy and hardware performance of inference engines. In this work, we further investigated the impact of the "analog" emerging non-volatile memory (eNVM)'s non-ideal device properties for on-chip training. By introducing the nonlinearity, asymmetry, device-to-device and cycle-to-cycle variation of weight update into the python wrapper, and peripheral circuits for error/weight gradient computation in NeuroSim core, we benchmarked CIM accelerators based on state-of-the-art SRAM and eNVM devices for VGG-8 on CIFAR-10 dataset, revealing the crucial specs of synaptic devices for on-chip training. The proposed DNN+NeuroSim V2.0 framework is available on GitHub at https://github.com/neurosim/DNN_NeuroSim_V2.0.

## I. INTRODUCTION

The state-of-the-art deep convolutional neural networks (CNNs) have shown remarkable breakthroughs in various applications, including speech recognition and image classification. As the popular CNNs tend to introduce huge amount of high-dimensional convolutional layers and hundreds of megabytes of parameters, to solve the bottleneck of extensive data transfer in the conventional von Neumann architectures, compute-in-memory (CIM) has emerged as a promising paradigm for designing the machine learning hardware accelerator [1].

Emerging eNVM devices such as RRAM [2], PCM [3], EpiRAM [4], ECRAM [5] and FeFET [6] have been proposed by the device community as candidates of "analog" synaptic devices, to represent the weights of deep CNNs in CIM accelerators. To evaluate these device properties from system-level perspective, we published a prior work in IEDM 2019 [7], whose latest version is named as DNN+NeuroSim V1.1, and served as an end-to-end benchmarking framework for the inference engine design. It supports flexible deep CNN topologies and versatile device technologies from CMOS to beyond-CMOS, with automatic CIM floorplanning to evaluate inference engines hierarchically. We focused on the impact of variability/reliability in synaptic devices, such as conductance variation and retention, and analog-to-digital converter (ADC) quantization loss, to investigate the trade-offs among inference accuracy, energy efficiency, throughput, chip area and memory utilization. Therefore, the DNN+NeuroSim V1.1 can be used as a supporting tool to find optimal design options of CIM inference engines for various synaptic device candidates and neural networks. By benchmarking the popular synaptic devices (including RRAM [2], and FeFET [6]) on CIM inference engine for VGG-8 with CIFAR-10 dataset, we learnt that, for "analog" synaptic device based CIM designs, the parallel read-out scheme and large on-state resistance (>100kΩ) are two of the most important specs to achieve superior energy efficiency (in TOPS/W) and throughput (in TOPS) in the CIM accelerators for inference.

To further study the potential applications of various synaptic devices for on-chip training, we introduce more non-ideal properties of synaptic devices that are critical for in-situ training accuracy, such as nonlinearity and asymmetry, device-to-device and cycle-to-cycle variation during weight update in Fig.

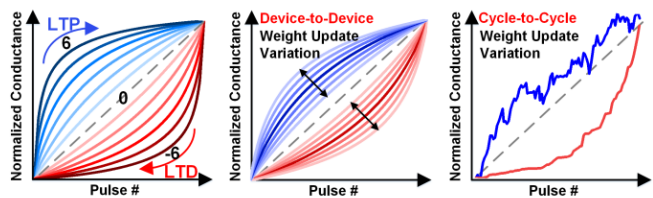

Fig. 1. Non-ideal properties on synaptic devices for in-situ training, including nonlinearity and asymmetry, device-to-device and cycle-to-cycle variation during weight update.

1 [8]. In CIM accelerators, to support on-chip training, we also implement extra peripheral circuits to calculate error and weight gradient in back-propagation.

In this work, we extended DNN+NeuroSim framework to support evaluation of training performance in CIM accelerators, and benchmark across SRAM and versatile eNVM devices for VGG-8 on CIFAR-10 dataset. The rest of the paper is organized as follows: ***Section II*** introduces the framework structure of DNN+NeuroSim V2.0. ***Section III*** describes the detailed architectures to support feed-forward and back-propagation computation in deep CNNs. ***Section IV*** discusses about benchmark results of CIM accelerators in on-chip training with versatile synaptic devices. ***Section V*** summarizes the work.

## II. INTEGRATED FRAMEWORK PRINCIPLES

Fig. 2 shows the framework structure of DNN+NeuroSim V2.0. As what has been proposed in prior framework V1.0 [7], the NeuroSim core is wrapped by python library, to support flexible network topologies, the default model is VGG-8 for CIFAR-10 based on low precision training method WAGE [9]. However, larger model such as ResNet-18 is also supported and arbitrary CNN topology could be defined by the user.

As shown in Fig. 2 (b), during training phase, non-ideal properties of synaptic devices during weight update are introduced into the python wrapper, including nonlinearity and asymmetry, device-to-device and cycle-to-cycle variation. It should be noted that, the number of pulses that will be applied to each synaptic device (to update the weights) is defined in a linear relationship with the calculated weight gradients (by digital circuit modules), however, due to the nonlinear and asymmetric model, the actual updated weights is not simply by adding linear weight gradients anymore, with further device-to-device and cycle-to-cycle variation, the non-ideal weight update will lead to accuracy degradations in in-situ training.

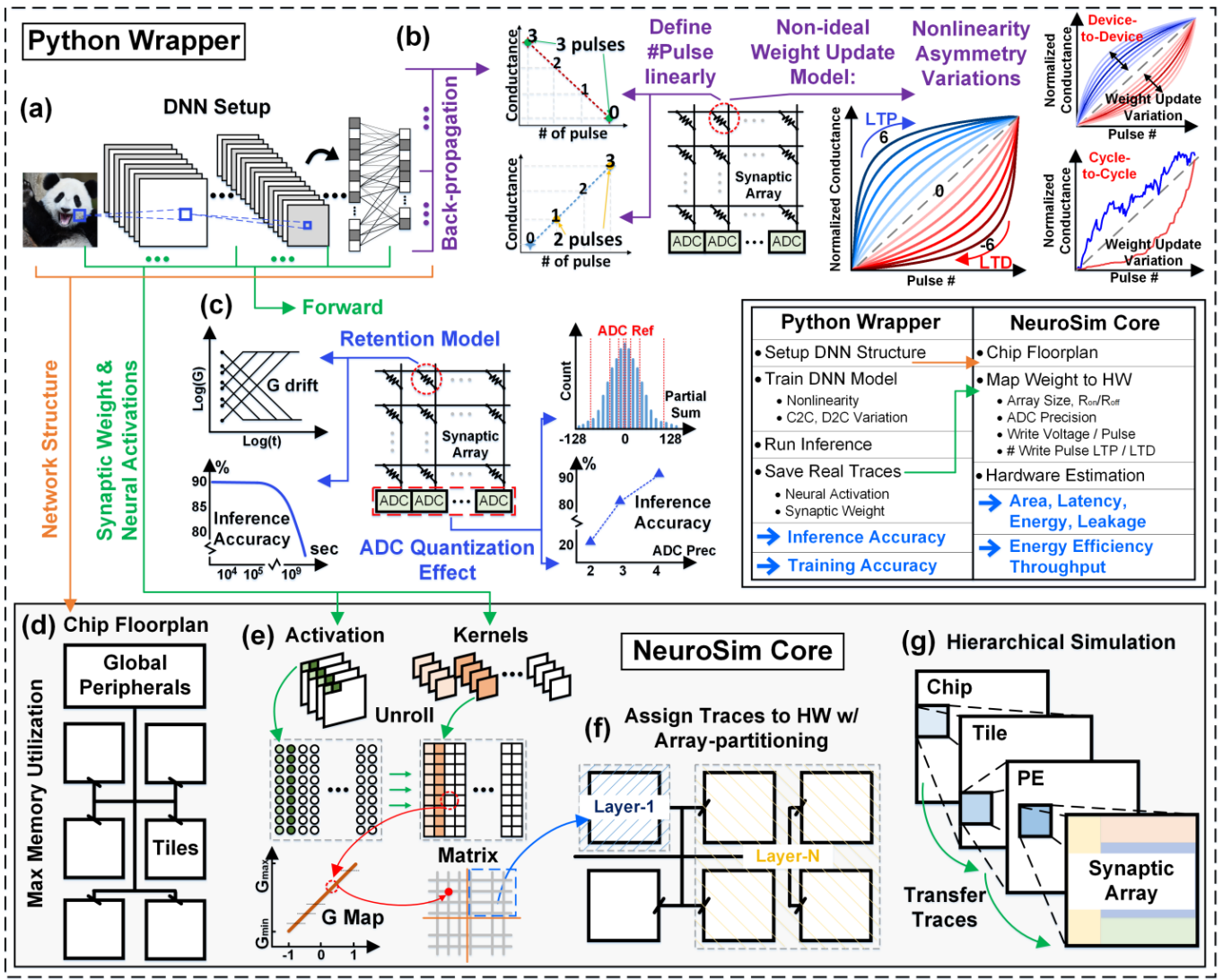

Fig. 2. Framework structure of DNN+NeuroSim V2.0. (a) DNN setup in python wrapper; (b) during training, introducing non-ideal properties of synaptic devices, including nonlinearity, asymmetry, device-to-device, and cycle-to-cycle variation during weight update; (c) during inference, introducing retention model and ADC quantization effects; (d) pre-defined network structure is loaded as input to NeuroSim core, for automatic floorplanning with weight-duplication to maximize memory utilization; (e) loading real trace (synaptic weights and neural activations) into NeuroSim, mapping data to conductance and digital voltage input cycles; (f) to be partitioned and assigned to different locations of the CIM system; (g) hierarchical simulation from chip to tile, and from processing element (PE) to synaptic array.

In Fig. 2 (c), we show the device retention model [10] and ADC quantization loss of partial sums during feed-forward operations, which was previously introduced in V1.0 [7]. Fig. 2 (d) shows the simulator taking network topology as input to automatically design the chip floorplan, while weight-duplication [11] is introduced to maximize memory utilization (defined as percentage of the used memory over the total memory) to certain layers in the network. This is a feature needed for convolutional layer where the unrolled kernel size is smaller than the memory sub-array size, in order to speed up DNN processing.

Fig. 2 (e) shows that in python wrapper, the neural activations and updated synaptic weights are stored for feed-forward evaluation, while the old synaptic weights (before weight update) are also stored for weight update evaluation during back-propagation. Within one epoch, due to the different weight gradients in each iteration, the hardware performance will also be different, however, to limit the overhead of running time in the framework, by default, we only take the real traces (neural activation, new and old synaptic weights) from the last iteration in each epoch, and run the traced-based simulation in NeuroSim core only once for every epoch. By doing so, we could guarantee reasonable simulation time, while still take a track of the hardware performance among different epochs during training.

Fig. 2 (f) shows the traces are partitioned and assigned to different locations of the chip according to the automatic floorplanning rule as introduced in V1.0 [7]. The top-down hierarchy of the CIM system is defined as chip, tile, processing element (PE) and synaptic array. The framework outputs include the hardware-constrained training or inference accuracy (from python wrapper), and hardware metrics such as chip area, latency, dynamic energy, leakage power, as well as energy efficiency and throughput (from NeuroSim core) for training or inference. The modular circuit component estimation are all calibrated by SPICE simulations across technology nodes from 130nm down to 7nm with PTM models [12], as shown in our prior work [13].

## III. CIM Architecture for Training

In CIM accelerators, to support training, additional peripheral circuits to calculate error and weight gradient are necessary to be implemented. In this section, we discuss about the detailed architectures for the four key steps in training, namely, 1) feed-forward, 2) computation of error, 3) computation of weight gradient, and 4) weight update. We firstly introduce the main hierarchical design of the entire CIM architecture, and then breakdown to the details of different computation steps.

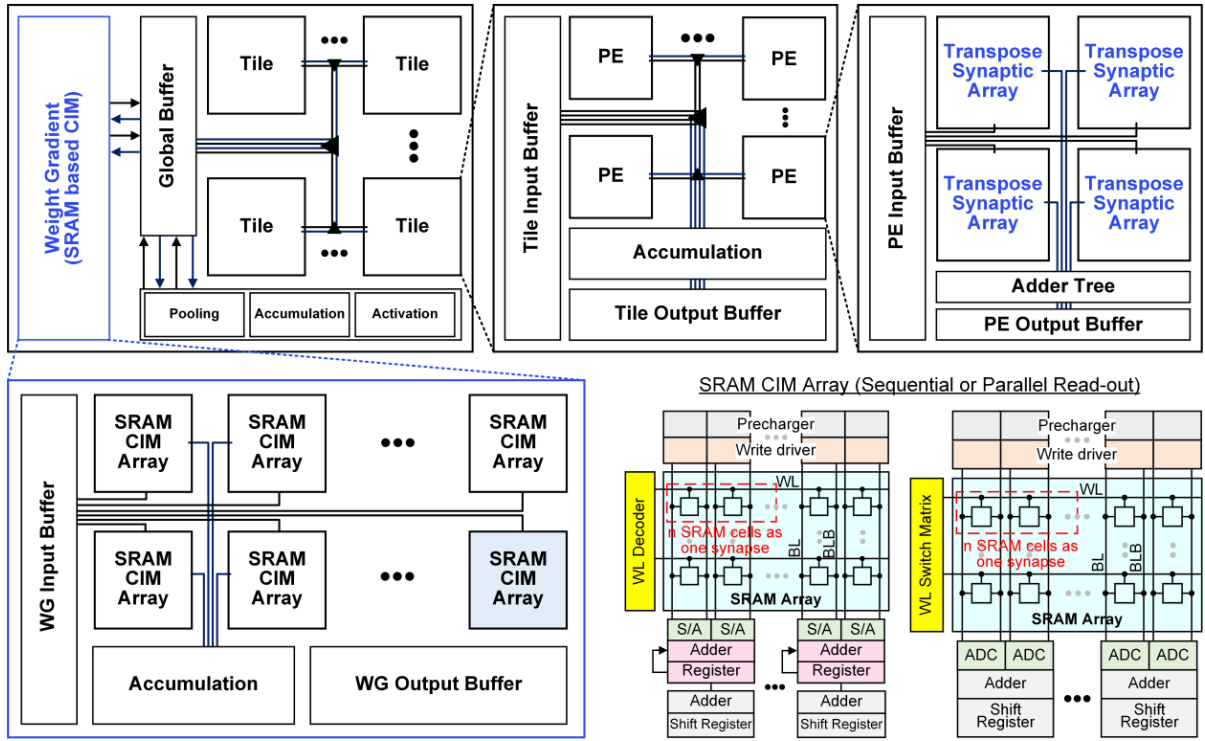

Fig. 3. Architecture structure defined in the simulator, the top level of chip contains tiles, global buffer, neural-functional peripheries (including pooling, accumulation and activations) and weight gradient computation function which is mainly built up with a group of SRAM-based non-transposable CIM arrays. Inside a tile, it is further portioned into multiple processing elements (PEs), while each PE consists of several synaptic arrays, along with adder trees and local buffers. H-tree routing is used for interconnect. To support training, the synaptic arrays are designed transposable.

## A. CIM Architecture

As Fig. 3 shows, in chip-level, the key components of the CIM accelerators are tiles, global buffer, neural functional units including pooling, accumulation and activation. There are four main hierarchies in the CIM accelerators, chip level, tile level, processing element level and synaptic array level. In different levels, peripheries are introduced, including buffers, interconnects (based on H-tree routing), and computational units (such as adder trees).

Besides these, to support training, we also implement the weight gradient computation function, which is mainly built up by a group of SRAM-based CIM arrays, local buffers and accumulation units. The reason that we choose the SRAM-based CIM arrays is, we need to frequently write data into the array and do vector-matrix multiplication. Although SRAM is not as area-efficient as eNVMs, and also has the problem of standby leaking, its fast writing performance still makes it more suitable and infinite endurance for the gradient computation compared with the eNVMs. More details of the weight gradient computation function will be discussed in *Section III-D*.

Moreover, since normally mini-batch based training with batch size=**B** is used, it means the number of intermediate data that need to be utilized and stored is huge. For example, during feed-forward, the **B** activations of all the layers will be stored to be used for the computation of weight gradients later in backpropagation; the **B** computation of errors of all the layers obtained in backpropagation will be stored; and the **B** weight gradients in one batch will be stored and accumulated to produce the delta weight for the final weight-update. Therefore, if the batch size is **B**, we have to store **B** copies of activations, errors and weight gradients of the entire network, before we can update the weights for a specific batch. To limit the on-chip buffer overhead, we assume that those data will be sent back to off-chip memory (i.e. DRAM) for the entire batch, and will be retrieved back to chip for error and weight gradient computation (*Section III-C* and *III-D*).

Due to this, and also to simplify the operation strategy, we assume that: the feed-forward, computation of error, and computation of weight gradient across the batch and weight-update will not be operated simultaneously on the CIM accelerators, which helps us to limit the hardware overhead for these additional back-propagation computational units (*Section III-D*). As Fig. 4 shows, in feed-forward (# 1), as batch size is assumed as **B**, the **B**× images will be fetched to on-chip global buffers one by one from the off-chip memory, and then to the CIM arrays for computation; meanwhile, the activations of each layer will be sent back to off-chip memory for the **B**× images. After that, the **B**× errors will be sent to on-chip global buffers one by one from the off-chip memory (# 2), and to CIM arrays for computation; while the errors of each layer will be sent back to off-chip memory for the whole batch.

Similarly, during the computation of weight gradients (# 3), **B**× errors and **B**× activations will be sent to on-chip buffers from the off-chip memory, and to the weight gradient computation units. As the errors will be reused (i.e. if kernel size is **K×K**, the activations will slide over the errors by **K×K** times to generate the weight gradients at **K×K** spatial locations), we will map the errors into the SRAM-based CIM arrays, and the activations will be applied as the input vectors. It should be noted that, as the SRAM-based CIM architecture is not area-efficient, we will not implement large amount of SRAM-based CIM arrays to support the weight gradient computation for all the layers. On the contrast, we assume the SRAM-based CIM arrays are large enough to support at least the largest layer in the whole network. Hence, during (# 3) in Fig. 4, the errors will be fetched to on-chip buffers layer by layer for each image, and image by image for the whole batch. For the layers whose unrolled errors' sizes are smaller than the SRAM-based CIM arrays, we will duplicate the errors in SRAM-based CIM arrays to speed up the computation. During this process, the weight gradients will be sent back to off-chip memory layer by layer for each image, and image by image for the whole batch.

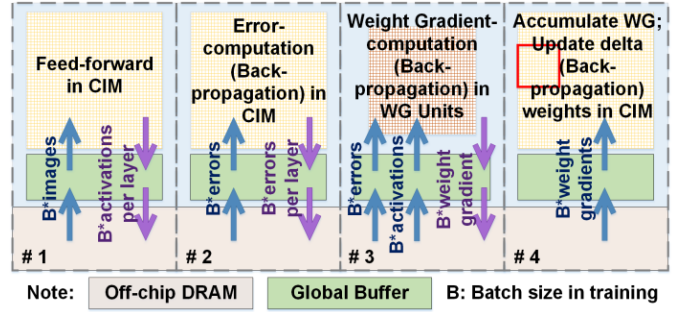

Fig. 4. The schedule of mini-batch based (batch size is B) training in CIM architecture.

Finally, to update the weights, we need to accumulate the weight gradients and calculate the delta weights. Since the precision of weight gradients is normally quite high, the requirement of on-chip hardware to accumulate and store **B**× weight gradients is pretty high (even for the smallest layer). In this case, we assume that, the on-chip accumulation units (precision and numbers) should at least support **B**× weight gradients with size equal to one CIM synaptic array (in DNN+NeuroSim V2.0, we assume memory cell precision to be equal to synaptic weight precision). As Fig. 4 (# 4) shows, to calculate the delta weight, each layer will be partitioned into multiple parts, and for each parts, the gradients will be sent to on-chip buffer and accumulated image by image, after the weight gradients are accumulated for entire batch, one specific CIM synaptic array will be updated. Meanwhile, we can start with the computation of next part of current layer. Thus, the weight gradients will be fetch to on-chip part by part (for the whole batch) and layer by layer, i.e. the CIM synaptic arrays will also be updated one by one for each layer, and layer by layer for the whole network.

In DNN+NeuroSim V2.0, we do not consider pipeline among the four key steps in training, i.e. #1 feed-forward, #2 computation of error, #3 computation of weight gradient, and #4 weight update, but the users can potentially optimize the design as done in other works [14][15]. However, the framework provides an option to build up pipeline system for feed-forward and computation of error, as we assume all the weights are stored on-chip in CIM synaptic arrays, we can process multiple images simultaneously on-chip, i.e. inter-image pipeline with on-chip global buffer overhead. This assumption (of four separated training steps) helps us to limit the global buffer size, as the global buffer will not be used to support different operations (feed-forward and back-propagation), the specs for global buffer is: what is enough to hold all the intermediate data to complete the computation for each layer or each step as mentioned above. In other words, the maximum requirements of on-chip buffer for the four separate operations (feed-forward, computation of error, computation of weight gradient across the batch and weight-update) will decide the size of global buffer. For example, in a simplest design, where the operations of #1 feed-forward and #2 computation of error are layer by layer, and image by image, while #3 computation of weight gradient and #4 weight update are assumed for one (synaptic CIM) array by array, and layer by layer, the global buffer size will equal to MAX(activation size of largest layer, error size of largest layer, weight gradient size of largest layer). Similarly, the on-chip computational hardware (such as accumulation units) are also limited, thus, the overall on-chip buffer size of the CIM accelerator is still acceptable even for training.

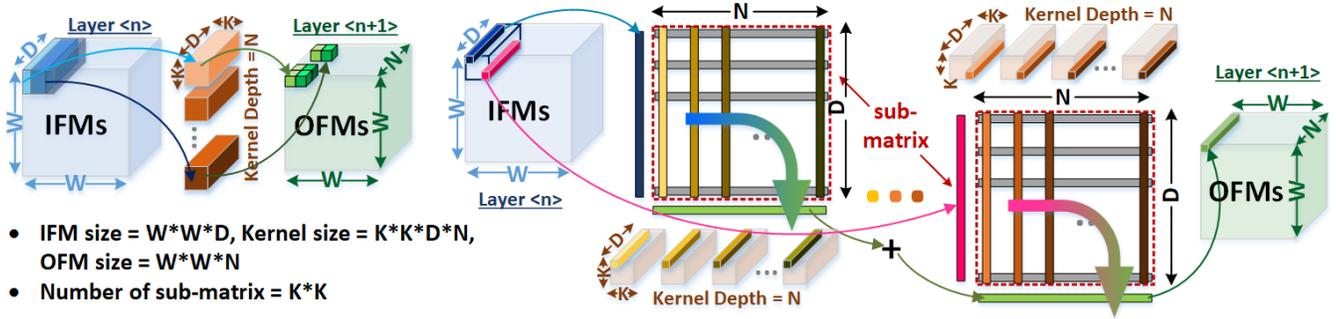

Fig. 5. A novel mapping method of convolutional layer in feed-forward [16] to maximize input data reuse, where the weights are mapped along their spatial location to a group of sub-matrix. K×K kernel is mapped to K×K sub-matrices (or processing elements, PEs). Partial sums are accumulated by adder tree.

### B. Feed-Forward

In Fig. 5, the computation of convolutional layer during feed-forward is shown as computation among tensors. In layer<n>, the size of input feature maps (IFMs) is **W×W×D** (where **D** is the depth of input feature channel), which are the outputs from layer<n-1>. The size of each 3D kernel is **K×K×D**, with kernel depth of **N** (i.e. there are **N** such 3D kernels), thus the total size of the kernels in layer<n> will be **K×K×D×N**. To get the outputs, a group of IFMs (with size **K×K×D**) will be selected at each time, and to be multiplied and accumulated with N kernels with size **K×K×D**, then each of them will generate a 1×1×1 output, the output from the top kernel (shown as light orange cube) goes to the front, and the output from the bottom kernel (shown as dark orange cube) goes to the back, thus, in total there will be 1×1×**N** outputs. It could be considered that, the kernels are "sliding over" the IFMs, and perform element-wise multiplications with a certain stride, and then the products of each element-wise multiplications in each 3D kernel will be summed up to get the final outputs, it is easy to detect that during the "kernel sliding", part of the input data will be reused for the computation of the next output. If we consider same-padding of the IFMs with a stride equals to one, it is straightforward to know that the output feature maps (OFMs) of layer<n+1> will be **W×W×N**, here the **N** (kernel depth) defines the depth of output feature channel.

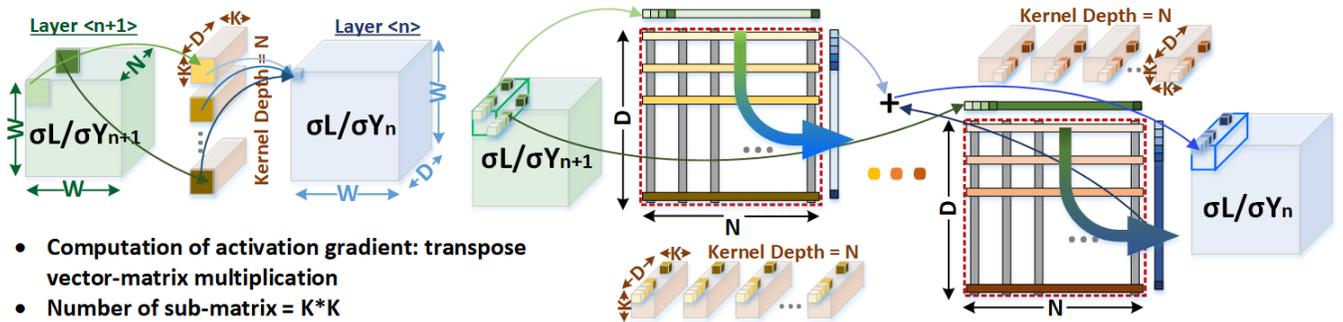

Fig. 6. The computation of error in CIM during backpropagation, based on novel mapping method [16] of convolutional layer.

To realize the input data reuse practically, we have proposed a novel mapping method and data flow for CIM inference in prior work [16], where the weights at different spatial location of each kernel are mapped into different sub-matrices. As Fig. 5 shows, if we cut each **K×K×D** kernel along its first and second dimension, we will get several 1×1×**D** partitioned kernel data, and for each kernel, there are **K×K** of them. According to the spatial location of partitioned data in each kernel, we define which group of these partitioned data should belong to. For example, all the partitioned data who are locating at the left-most and top-most position at each kernel, will be considered as one group, and implemented into one sub-matrix, the height and width of each sub-matrix should equal to 1×1×**D** and **N**. Hence, **K×K** sub-matrixes are needed for the kernels (who's first and second dimension equal to **K** and **K**), since each sub-matrix has size **D×N**, and the size of total weight matrix is **K×K×D×N**. Similarly, the input data which should be assigned to various spatial location in each kernel, will be sent to the corresponding sub-matrix respectively. To practically map and operate large convolutional layers on chip, array partitioning [17] is introduced, which could cut a single large matrix into several sub-arrays, and parallelize the computation efficiently.

In this framework, we also utilize this novel mapping method for CIM training, since it groups the kernels according to their special location, which makes it easier to implement transposable synaptic array to calculate the errors efficiently.

## C. Backpropagation for Error

As Fig. 6 shows, during back-propagation, to calculate the errors, the errors (i.e. the gradient of loss function respective to the activation) from deeper layer need to be fetched backwards and do the element-wise multiplication and accumulation with the prior kernels. In layer<n>, the error from layer<n+1> will be the input data, and be separated into different channels, then applied to corresponding kernels. For example, the error from first channel (shown as light green plane) will be fetched to the first kernel (shown as light yellow plane) to do the element-wise multiplication, and the accumulated output will be the first element in the first channel of layer<n>'s error. According to the novel mapping method, the kernels are partitioned based on their spatial location and collected along their channel (dimension=**D**), and mapped into the columns in CIM arrays. It would be considered that, the rows in such CIM arrays are the weights in a specific location in each kernel from different channels. For example, as shown in Fig. 6, the first-channel weights at left-most and top-most location of each kernel (shown as light yellow nodes) are mapped as the first row in the first sub-matrix; the last-channel weights at right-most and bottom-most location of each kernel (shown as dark orange nodes) are mapped as the last row in the last sub-matrix.

In this case, since the results at the same channel in different kernels will be accumulated to get the outputs, it is straightforward to find that, we can automatically accumulate the products along each rows in the CIM array and sum up the partial-sums among sub-matrix to get the error. Thus, the error in layer<n+1> will be sent to each sub-matrix according to their spatial locations during kernel-sliding, and each of them will be considered as inputs to each column in the CIM array.

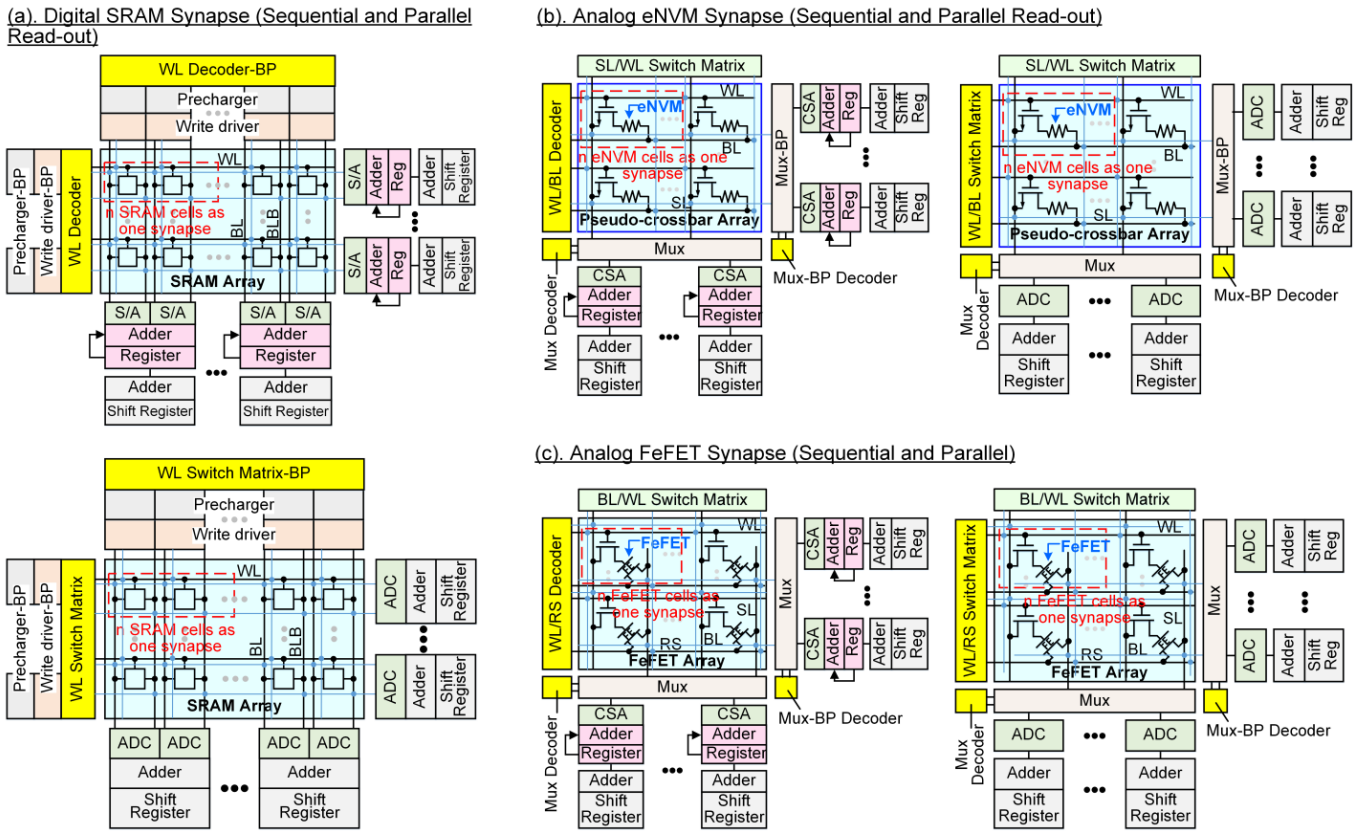

Fig. 7. In DNN+NeuroSim V2.0, transposable synaptic arrays are implemented to support on-chip training. Available synaptic devices are (a) SRAM, (b) two-terminal eNVMs (e.g. RRAM) and (c) three-terminal eNVMs (e.g. FeFET), with both sequential and parallel read-out schemes.

Therefore, we need to modify the design of synaptic arrays in the CIM architecture, to support both conventional (feed-forward) and transposed (error) computations. As shown in Fig. 7, in this framework, we provide the transposable synaptic array designs with versatile synaptic devices, ranging from SRAM, two-terminal eNVMs (like RRAM) and three-terminal eNVMs (like FeFET), and can be designed with either sequential or parallel read-out schemes.

In SRAM-based synaptic arrays, during feed-forward, the conventional computation scheme is to activate each rows, then read out and accumulate the products along the columns. During back-propagation, to calculate the errors, we need to activate each columns and accumulate the product along rows. To do so, we need to implement transposed word-lines, bit-lines with 8T-SRAM cells [18] (duplicate and rotate 90 degree from the original ones, shown as light blue lines) to realize transposed computation, with additional peripheral circuits (such as WL decoder or switch matrix, sense amplifier or ADC, and shift-adder with registers).

However, in eNVM-based synaptic arrays, since there are SL switch matrix (in two-terminal eNVM designs) or BL switch matrix (in three-terminal eNVM designs) for weight update, in transposed computation (for error), we could use the SL or BL switch matrix to rotate the input and the output. In this case, for eNVM-based synaptic arrays, we only need to add sense amplifier or ADC (along with adder, shift-adder and registers) to read out the partial-sums horizontally along rows.

In this framework, as we assume that the feed-forward and back-propagation will not be operated simultaneously, we could avoid complex circuit designs (for logical control) above the synaptic array level.

*D. Weight Gradient Caculation*

When the computation of error is done and stored to the off-chip DRAM [19] memory (for each batch), we need to start the computation of weight gradient. As Fig. 8 shows, to calculate the weight gradients of layer<n>, the error from layer<n+1> will be applied to do element-wise multiplication and accumulation with the activations from layer<n> in a channel-to-channel scheme.

For example, a part of the first-channel activations in layer<n> (shown as light green plane) will be multiplied with the error in layer<n+1> (shown as light blue plane), the element-wise products will be accumulated and be the first weight gradient in the first channel of the first kernel (shown as light yellow node). Similarly, the last-channel activations of layer<n> (shown as dark green plan) will be multiplied with the last-channel error of layer<n+1> (shown as dark blue plane), and accumulated to be the first weight gradient in the last channel of the last kernel (shown as dark orange node). During this process, we can get the weight gradients at left-most and top-most spatial locations through all the channels for each kernel. To get all the weight gradients, we need to sweep the activations and repeat the multiplication-and-accumulation with the error by **K×K** times, representing **K×K** spatial locations in the kernels.

Therefore, we could easily unroll each channel of the error into a long column, as the products will be accumulated inside each channel, and with number of channels equals to **D**, there will be **D** such long columns to form a large matrix. The activations will be treated as the inputs to the matrix, with channel depth equals to **N**, there will be **N** unrolled vectors of inputs applied to the matrix, to get one group of the weight gradients (corresponding to the weights at a specific spatial location of each kernel, and also representing one of the **K×K** sub-matrix in novel mapping of the weights). Thus, to sweep the activations by **K×K** times to get all the weight gradients, in total there will be **K×K×N** unrolled vectors of activations applied to the matrix of errors. Since the dimension of convolutional layers in the popular deep CNNs could be quite large, yielding a large matrix to store the errors in each layer, we will also introduce array-partitioning [17] into this weight gradient computation, to avoid large memory operations.

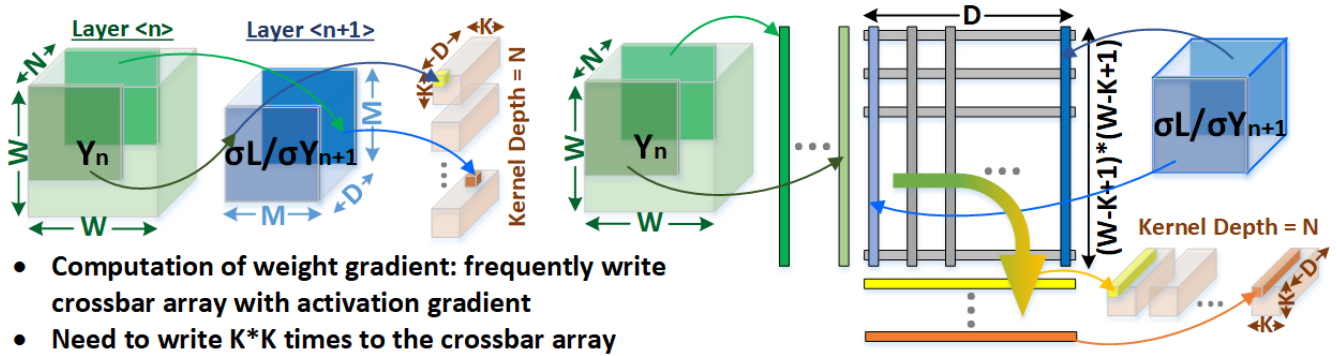

Fig. 8. The computation of weight gradient in CIM.

Since the weight gradients need to be calculated for various inputs across the batch, and finally accumulated to get the delta weights for each iteration, we can find that, we need to frequently re-write the matrix of errors for different images, which leads to a huge overhead of memory writing latency and energy. Due to this, we choose the SRAM-based CIM non-transposable array for weight gradient calculation (as Fig. 3 shows) over other eNVM-based designs, to avoid the huge memory programming overhead and the limited endurance. It is noticeable that, the SRAM-based CIM design could cause a larger area overhead compared with eNVM-based ones. To minimize the on-chip hardware resources for weight gradient computation, we do not need to process all the layers simultaneously, but just perform layer-by-layer weight gradient computation. Therefore, we can only put enough SRAM-based CIM arrays to support the layer with largest size of errors through the entire network, and hence, thus we find that the area overhead of SRAM-based weight gradient computation is acceptable.

*E. Weight Update*

During the computation of weight gradients (across the batch), the weight gradients of each image will be sent back to the off-chip memory successively. Therefore, before we can update the weight in the CIM accelerator, we need to load the weight gradients back and accumulate them to get the delta weights of each layer. Similarly as what we designed for the weight gradient computational units, to minimize the area overhead, we assume that as long as the accumulation units in chip-level can support one specific portion of weight gradients accumulation (across the batch), it is acceptable for us to process the weight update.

For example, we design the chip-level accumulation units which are enough to support the weight gradients accumulation (across one batch) with size equals to the synaptic array size. For a specific layer, we firstly choose to update the first synaptic array (one portion of the weight matrix in this layer), so we just load in the corresponding parts of weight gradients for this specific layer across the batch and do the accumulation. The delta weights for this portion of the specific layer will be stored into the global buffer, and then transfer to the exact synaptic array for updating.

For example, as Fig. 9 shows, we are processing the weight gradient accumulation and weight update for a specific synaptic array in layer<i>. At the very beginning (T=1), we preload the weight gradients of the first two image in the batch (to global buffer), at next cycle (T=2), these two weight gradients will be loaded to accumulation units for computation (green arrows), meanwhile, the weight gradients of the third image will be loaded into the global buffer (red arrow), and the accumulated gradients will be sent back to the global buffer (blue arrow). Similarly, we can continue these operations for the following images, until we accumulate the gradients for the whole batch. It is clear to find that, as we assumed enough accumulation units to support accumulation of weight gradients (with size equals to one synaptic array), at a specific cycle, the new gradients and the accumulated gradients (from prior cycle) can be loaded to accumulation units and be processed simultaneously, and generate same amount of accumulated gradients (to be saved back to global buffer). Thus, as long as the global buffer is released and the data are safely operated inside the accumulation units, we can start to load in new weight gradients, and also it is safe to store the newly accumulated gradients to the released global buffer. In other words, during such successive accumulation operation, we only need the global buffer storage to be (2×synaptic-array-size×highest-possible accumulated weight gradients precision).

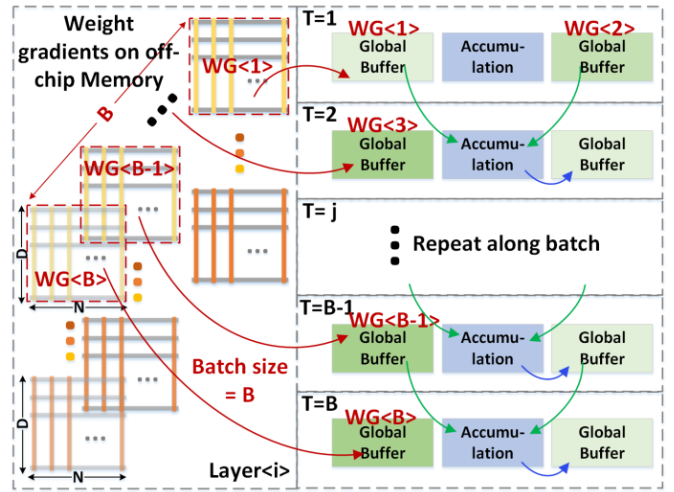

Fig. 9. Example of weight gradient accumulation.

Apparently, this extreme assumption can help us to significantly limit the area overhead for training, but as a trade-off, it will cause a large latency overhead during such successive weight gradient accumulation, i.e. B× (buffer-read + accumulation + buffer-write). Thus, in DNN+NeuroSim V2.0, we also provide an option to gradually increase global buffer size, namely, "buffer overhead constraint", with larger constraint ratio, the global buffer size increase, which can support weight gradients of more synaptic arrays, correspondingly, the chip-level accumulation units will also increase, and thus decrease the latency of gradient accumulation and weight update (more synaptic arrays can be updated simultaneously). With such option, the user can find an optimal design option with a specific area constraint.

## IV. BENCHMARK RESULTS

In CIM accelerators for inference accuracy and hardware performance, the most critical factors are the on-state resistance $R_{on}$ and ADC precision as we showed in our prior work [7]. In this work, we take more emphasis on the non-ideal synaptic device properties (including nonlinearity and asymmetry, device-to-device and cycle-to-cycle variation) for in-situ training. We benchmark across device technologies based on VGG-8 for CIFAR-10 dataset. To study the impact of device precision, we assume each eNVM based synaptic weight will be represented by only one device cell, with the exception that n-bit weight is represented by n SRAM cells.

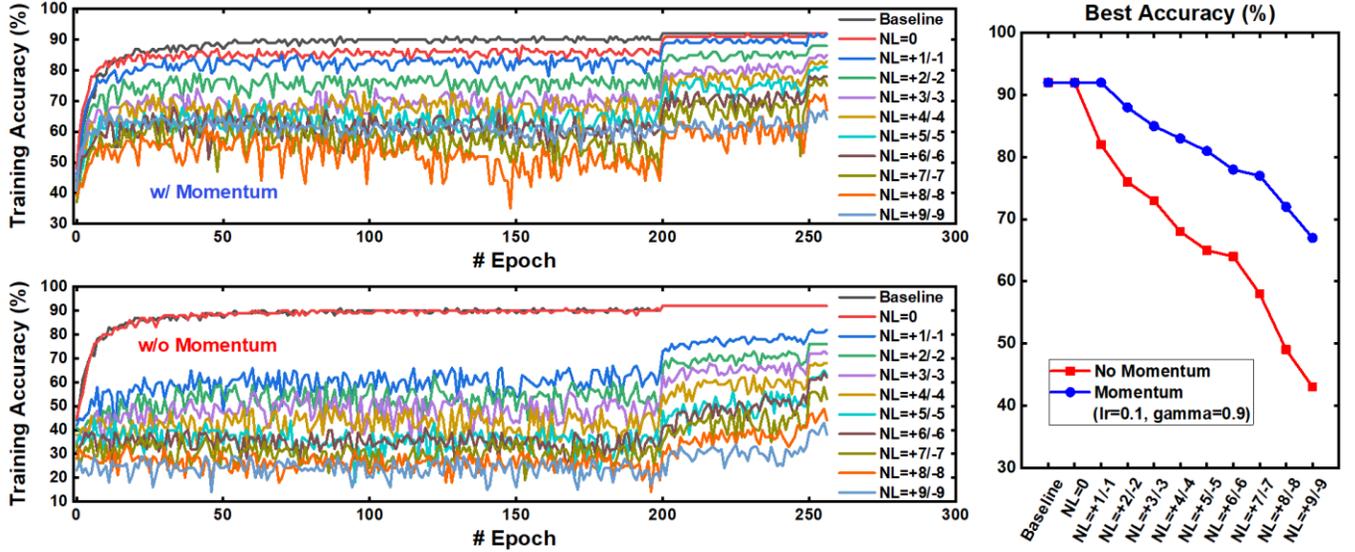

Fig. 10. Analysis of nonlinearity and asymmetry for in-situ training, w/ and w/o momentum optimization.

## A. Impacts of Non-ideal Synaptic Device Properties

To analyze the impacts of non-ideal synaptic device properties on the training accuracy, we introduced the models of nonlinearity and asymmetry, device-to-device and cycle-to-cycle variations (as shown in Fig. 1) into the python wrapper, to evaluate the degradation of training accuracy under these non-ideal properties. We run the VGG-8 on CIFAR-10 dataset, and sweep the different values for each non-ideal property, to quantify their effects individually. In this case, we fixed the precision of activation, weight, gradient and error to be 8-bit, which implies that the synaptic device precision is considered as 8-bit (256 levels).

The model of nonlinearity and asymmetry can be expressed by the following equations [6], where the updated conductance value is in a nonlinear relationship with the number of pulses ($P$), the $G_{max}$, $G_{min}$ and $P_{max}$ represent the maximum, minimum conductance values and the maximum number of pulses that synaptic device can achieve (i.e. pulse resolution). $A$ determines the nonlinear behavior of weight update, while $B$ is the function of $A$ that is adjustable along the range of $G_{max}$, $G_{min}$ and $P_{max}$.

$$G_{LTP} = B\left(1 - e^{-P/A}\right) + G_{min}$$

$$G_{LTD} = -B\left(1 - e^{(P-P_{max})/A}\right) + G_{max}$$

$$B = (G_{max} - G_{min})/(1 - e^{-P_{max}/A})$$

Fig. 10 shows the training accuracy under different nonlinearities with asymmetry, without device-to-device or cycle-to-cycle variations. A recent work [20] argues that the momentum optimization can significantly help to overcome the drawbacks of large nonlinearity and asymmetry, in this framework, we also introduced the momentum optimization method as below:

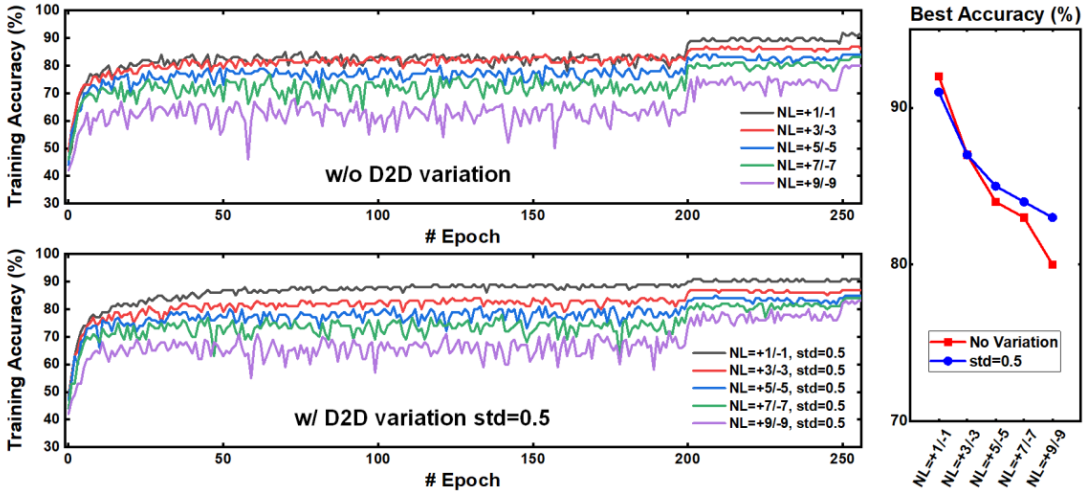

Fig. 11. Analysis of device-to-device variation under different nonlinearities, w/ momentum optimization.

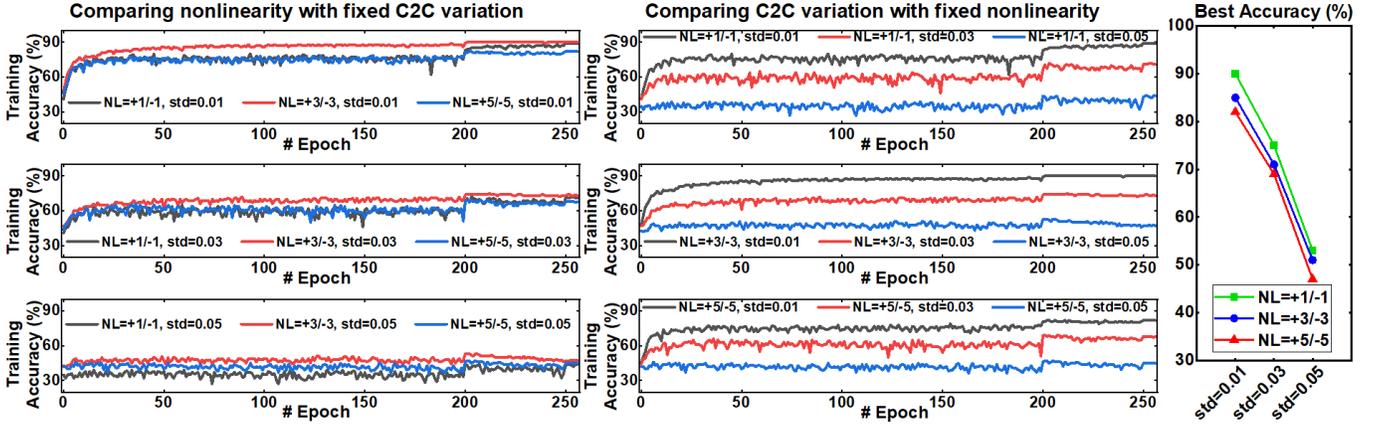
Fig. 12. Analysis of cycle-to-cycle variation under different nonlinearities, w/ momentum optimization.

$$v_j = \beta \times v_{j-1} + (1 - \beta) \times g_j$$
$$w_j \mathrel{-}= lr \times v_j$$

where $\beta$, $v_j$, $g_j$, $lr$ and $w_j$ donate the momentum, velocity, gradient and weight at j[th] batch of one epoch. In this work, we set momentum $\beta = 0.9$, and the results show that the in-situ training could achieve ~85% with a moderate asymmetric nonlinearity (NL=+3/-3) and ~77% even with a large asymmetric nonlinearity 6 (NL=+6/-6).

During weight update, the device-to-device (D2D) variation will introduce different nonlinearities to different synaptic devices. To build up the behavior model, we randomly generate the nonlinearities to different synaptic weights with a standard deviation (σ) respect to the mean nonlinearity value (μ). Fig. 11 shows the training with device-to-device variations under different nonlinearities (means), where the standard deviation is set to 0.5 (σ=0.5). The results show that, with momentum optimization, the device-to-device variation will not affect the accuracy, and when nonlinearity becomes larger, the device-to-device variation tends to salvage the accuracy degradation.

Furthermore, to study the impacts of cycle-to-cycle (C2C) variation, we integrate the behavior model in a similar way as the device-to-device variation. As the cycle-to-cycle variation is referred as the variation in conductance change at every programming pulse, in the proposed framework, we can express the cycle-to-cycle variation (σ) in terms of the percentage of entire weight range. Fig. 12 shows the results of training accuracy with different cycle-to-cycle variations (σ equals to 0.01, 0.03 and 0.05 respectively, implying 1%, 3% and 5% of the entire conductance range) under different nonlinearities (asymmetric NL equals to +1/-1, +3/-3 and +5/-5). With larger cycle-to-cycle variation, the accuracy drops significantly, this is due to the fact that: the cycle-to-cycle variation may cause a weight-update in an opposite direction to the desired one (ideal gradients), and thus induce opposite momentum directions.

### B. Hardware Performance Per Epoch

To analyze the hardware estimation, in DNN+NeuroSim V2.0, we produce detailed reports for each epoch in runtime, which include the area, latency and energy breakdown by main components, as well as total and peak latency and energy breakdown by operations (peak is defined as computation within synaptic array only, and do not consider operations related to off-chip memory, buffers or interconnect). Across the entire simulation, the reports of each epoch will be generated successively, meanwhile, several summarized reports will also be generated, which only contain main evaluation results, such as the accuracy, energy efficiency and throughput for each epoch, and the distribution (mean and standard deviation) of weights and delta weights of different layers in each epoch. The list of expected reports is shown below, as the default setting of this framework is on VGG-8 and CIFAR-10 dataset with 256 epochs for training, there will be 256 detailed reports in total:

> **Under path of DNN+NeuroSim V2.0 framework:**
> - NeuroSim_Results_Each_Epoch (folder):
>   - *Breakdown_Epoch_0.csv*
>   - ……
>   - *Breakdown_Epoch_256.csv*
> - *NeuroSim_Output.csv* **(summary of hardware performance)**
> - *PythonWrapper_Output.csv* **(summary of online learning accuracy)**
> - *Weight_dist.csv*
> - *Delta_dist.csv*
> - *Input_activity.csv*

To study the impact of device precision, during technological benchmark, we fix neural network structure, but will change the weight and gradient precision according to different device properties, i.e. we force the weight and gradient precision to be the same as eNVM device precision, because here we just use one single device to represent one synaptic weight. To avoid accuracy drop in ADC quantization, we assume the synaptic array to be 128×128 for all the designs, and use 6-bit ADCs with nonlinear quantization which guarantees high-enough accuracy even for the designs with 7-bit device precision.

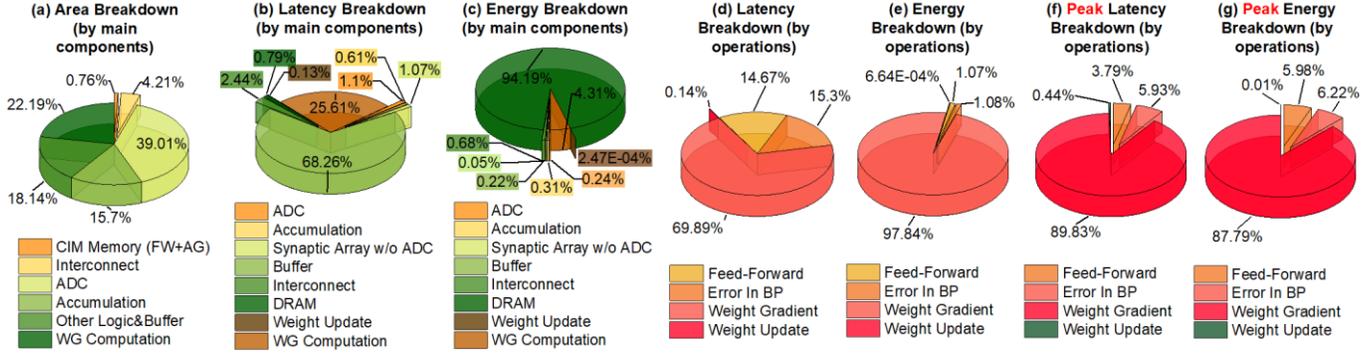

Fig. 13 In CIM accelerator for VGG-8 on CIFAR-10, DNN+NeuroSim V2.0 reports detailed hardware estimation results for each epoch. (a) area breakdown by main components; (b) latency and (c) energy breakdown by main components; (d) latency and (e) energy breakdown by operations; (f) peak latency and (g) peak energy breakdown by operations. The data shown is from the 100$^{th}$ epoch of FeFET-based [4] CIM architecture.

Fig. 13 shows the breakdown report in the 100$^{th}$ epoch of the FeFET-based [6] CIM accelerator. From Fig. 13 (a), we can find that the 6-bit ADC (flash ADC, i.e. multi-level sense amplifiers) is dominant in the total area, while the weight gradient computation units also occupy a lot of on-chip area, since they are built up by SRAM-based CIM arrays. Moreover, to support data transfer during each operation, the buffer and control circuits are also quite large. The area of accumulation includes chip-level accumulation units, tile- and PE- level adder trees, and adder or shift-adder in synaptic-arrays.

In Fig. 13 (b) and (c), we can find that, since there are a lot of data transfer on-chip, the buffer latency and DRAM energy are the bottleneck of the hardware performance. As Fig. 13 (d) and (e) show, we breakdown the total latency and energy into four main operations. The feed-forward and computation of error are quite similar, since their operation schemes, volume of input/output data and computations, and utilized hardware resources are quite similar (through the transposable CIM array). While the computation of weight gradients dominate in the total latency and energy, since during this operation, we need to load in the activations and errors from off-chip memory, and frequently write the SRAM-based CIM arrays for computation, then load out the weight gradients to off-chip memory. These repeated operations of off-chip memory access and SRAM write make the weight gradient computation to be the bottleneck in the entire training.

On the other hand, the latency and energy of weight update are ignorable, this is because we need to operate the prior operations for every input, but we only need to update the weights once per batch. In other words, the batch size is 200 in this benchmark, it means the latency and energy of weight update are averaged by 200× in each epoch, i.e. total latency and energy for each batch equals to (200 × feed-forward) + (200 × computation of errors) + (200 × computation of weight gradients) + (1 × weight update). In Fig. 13 (f) and (g), we can also find the peak latency and energy breakdown by operations. Similarly, the computation of weight gradients contributes most of the latency and energy, as it is based on SRAM and also includes repeated write operations.

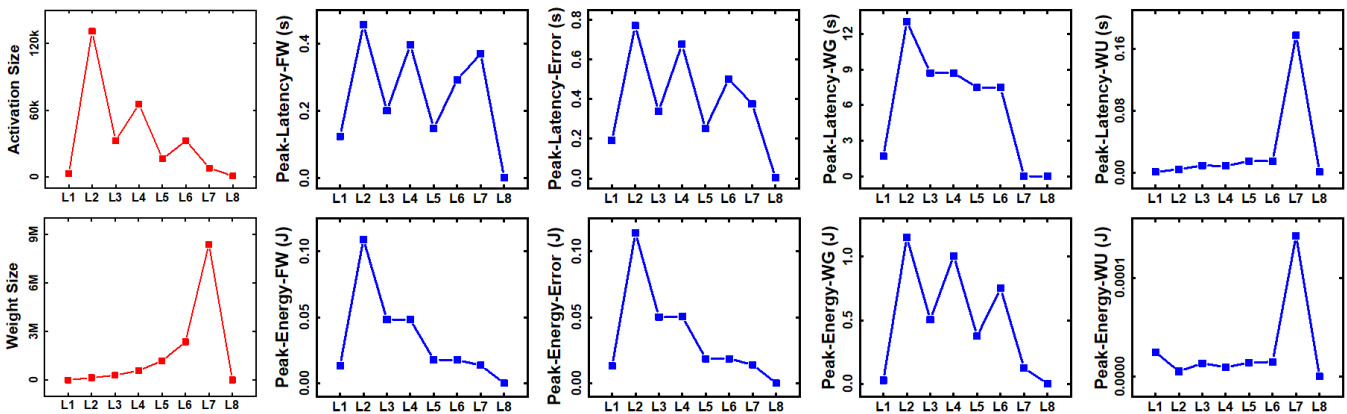

Fig. 14. Peak latency and energy across all the layers in VGG-8 of four main operations (feed-forward, computation of errors, computation of weight gradients and weight update) of one epoch. The data shown is from the 100$^{th}$ epoch of FeFET-based [6] CIM architecture.

In Fig. 14, the peak latency and energy of four main operations are shown across all the layers in the VGG-8 (blue lines and symbols), to analyze the relationship among layers, we also draw the size of activations and weights for each layer (red lines and symbols). In feed-forward and computation of errors, the latency and energy are in the same trend as the activation size (across the layers), which is reasonable since the activation size determines the number of operations in each layer. In the computation of weight gradients, the energy's trend fits with the activation size, while the latency trend is a little different, this is because in deeper layer, the activation size is smaller, so there are possibilities to compute more gradients simultaneously. As mentioned in *section III-A* and *III-D*, if we assume the SRAM-based CIM arrays are large enough to support the largest layer in the whole network, the layers whose unrolled errors' sizes are smaller than the SRAM-based CIM arrays, we will duplicate the errors in SRAM-based CIM arrays, fetch multiple activations simultaneously and speed up the computation. In DNN+NeuroSim V2.0, the NeuroSim core can automatically define the duplication scheme and take the "speed-up" into account. Finally, the latency and energy of weight update are only related to the weight size.

### C. Hardware Performance Across Epoch

To explore the hardware performance during the entire training process, we track the data (weight and gradient distributions) and the real-time estimations for every epoch. The feed forward and error computation employed the real-trace of activations and updated weights (every layer) for each epoch from python wrapper.

It should be noted that, to limit the simulation time in DNN+NeuroSim V2.0, we applied a "pseudo-traced" method to estimate the hardware performance of weight gradient computation. During estimation, we access to the binarized activations (quantized fix-point to digital format) and approximate the percentage of ones for each layer, which tends to represent the row-activities of the SRAM-based CIM arrays. Similarly, we can estimate the percentage of ones stored in the SRAM-based CIM arrays (as binarized errors). While for weight update, we access the old weights and updated weights (every layer) for each epoch, to process a real-traced estimation. In this way, we only need to access the activations, old weights and updated weights (every layer) at each epoch, to avoid huge data access in the framework during simulation, and still guarantee reliable estimation.

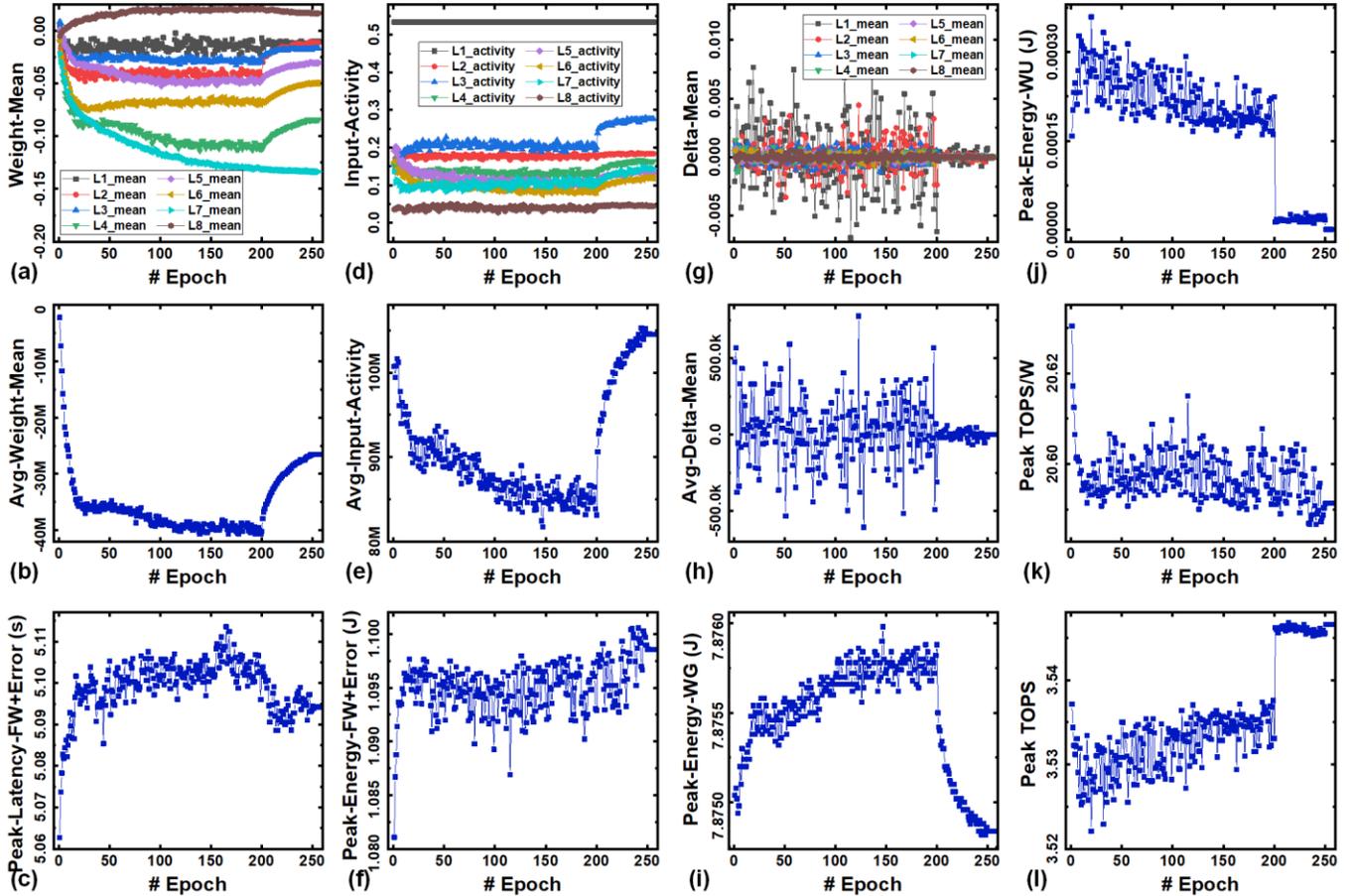

Fig. 15. (a) Weight means of each layer across all the epochs. (b) Weight means normalized with activation and weight size of each layer across the epochs. (c) Peak latency of feed-forward and error computation across epochs. (d) Input activities of each layer across all the epochs. (e) Input activities normalized with activation and weight size of each layer across the epochs. (f) Peak energy of feed-forward and error computation across epochs. (g) Delta weight means of each layer across all the epochs. (h) Delta weight means normalized with activation and weight size of each layer across the epochs. (i) Peak energy of weight gradient computation across epochs. (j) Peak energy of weight update across epochs. (k) Peak energy efficiency (TOPS/W) across epochs. (l) Peak throughput (TOPS) across epochs. The data shown is from the FeFET-based [4] CIM architecture.

As Fig. 15 shows, to study the hardware performance of each steps during training, we tracked the distribution of weights, delta weights and input activities across epochs, and find the relationship between traces and estimation results. The data is from the FeFET-based [6] CIM architecture design.

In Fig. 15 (a), we track the mean value of each layer's weight, during the entire training. Since the contribution of the weights from different layer is also determined by the weight size (defining amount of hardware resources) and activation size (defining number of computations), in Fig. 15 (b), we normalize the weight means of entire network, by multiplying the activation and weight size with the weight means of corresponding layer, and summing them up. Similarly, we track the input activities of each layer across epochs, and normalize according to the activation and weight size in Fig. 15 (d) and (e), as well as the delta weights in Fig. 15 (g) and (h). It should be noted that, the input activity of first layer does not change significantly since it is based on the image from CIFAR-10 dataset, while the activities of other layers increase after $200^{th}$ epoch (with tuned learning rate), because weight means increase after $200^{th}$ epoch (produce higher output feature maps). As for the delta weights, it is clear to see that the delta weight means are close to zero after $200^{th}$ epoch, since the network is converging. This also explains the trends of peak energy of weight gradient computation and weight update in Fig. 15 (i) and (j), as well as the significant increase of peak throughput after $200^{th}$ epoch in Fig. 15 (l).

We sum up the peak latency and energy of feed-forward and error computation as shown in Fig. 15 (c) and (f). In CIM architectures, the synaptic weights are mapped in a linear relationship with the conductance values. As the averagely contributed weight mean decreases with epoch, which means the overall conductance decreases, induces smaller currents

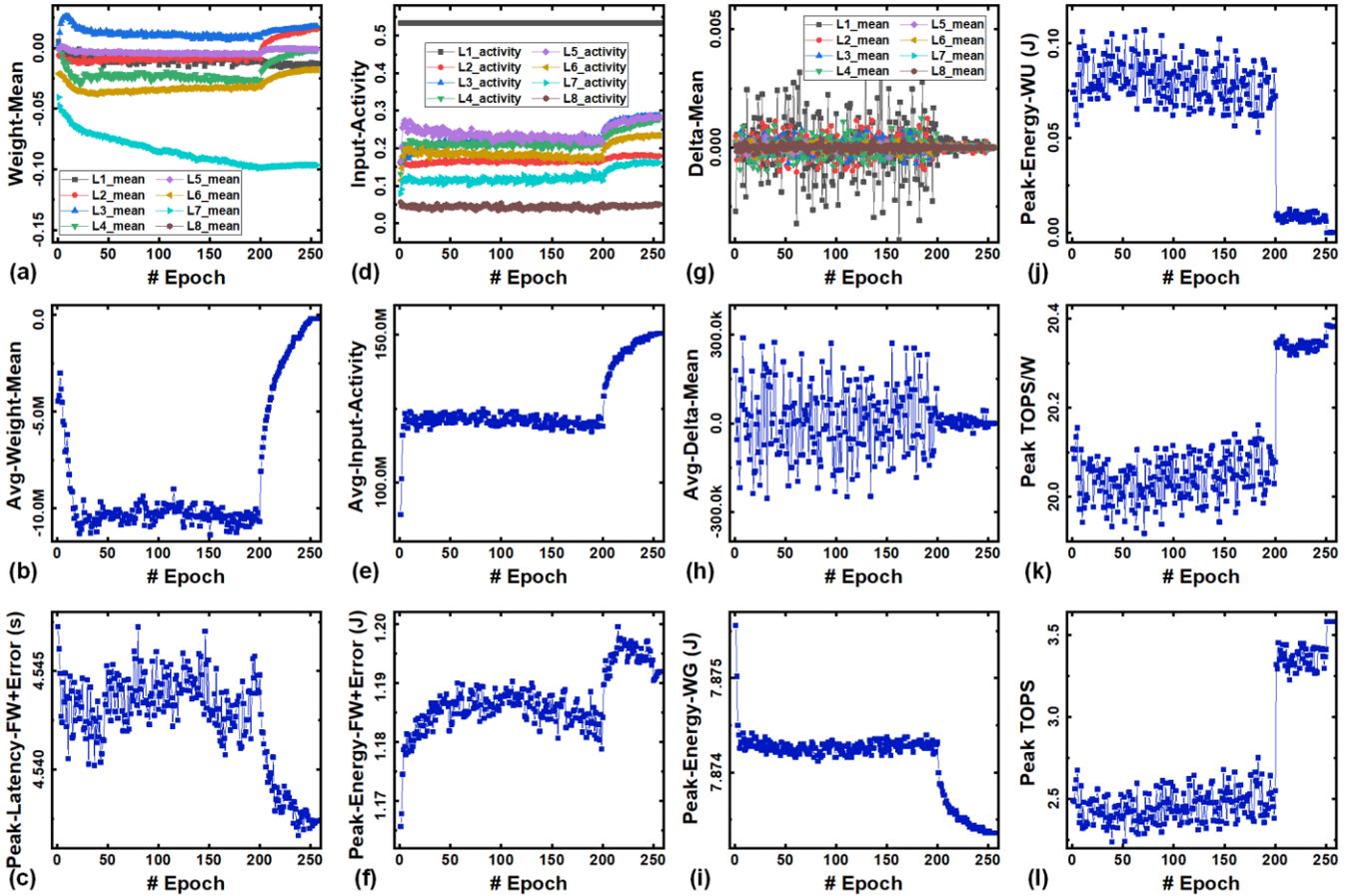

Fig. 16. (a) Weight means of each layer across all the epochs. (b) Weight means normalized with activation and weight size of each layer across the epochs. (c) Peak latency of feed-forward and error computation across epochs. (d) Input activities of each layer across all the epochs. (e) Input activities normalized with activation and weight size of each layer across the epochs. (f) Peak energy of feed-forward and error computation across epochs. (g) Delta weight means of each layer across all the epochs. (h) Delta weight means normalized with activation and weight size of each layer across the epochs. (i) Peak energy of weight gradient computation across epochs. (j) Peak energy of weight update across epochs. (k) Peak energy efficiency (TOPS/W) across epochs. (l) Peak throughput (TOPS) across epochs. The data shown is from the EpiRAM-based [4] CIM architecture.

along the synaptic arrays and cause longer latency (as Fig. 15 (c) shows).

As shown in Fig. 15 (b) and (e), before 200th epoch, both of the normalized weight means and input activities decrease, which denotes smaller column currents, and thus, we can find that in Fig. 15 (c) and (f), both of the latency and the energy in the synaptic CIM arrays (feed-forward and error computation) tend to increase with smaller conductance (larger resistance). This is because, with relatively high on-state resistance, the column currents are small (longer delay), while with larger resistance, the ADC delays tend to increase even more rapidly (although power decrease with smaller column currents), and overall caused the energy increase. This effect makes the energy of peripheral circuits less dominant, i.e. cannot see a clear trend of slightly decreased input activities in Fig. 15 (e) with affected energy in Fig. 15 (f), which is contrary to the EpiRAM [4] design (discuss later in Fig. 16). After 200th epoch, weight means and input activities increase dramatically, which leads to larger column currents and thus the latency decrease after 200th epoch in Fig. 15 (c). Meanwhile, although the latency will not be affected by the input activity itself (without column currents' effect) in parallel read-out scheme, there will be more transistors being activated simultaneously in the peripheral circuits (e.g. switch matrix), and thus, we can find the dynamic energy slightly increased with rapidly increased (much higher) input activities (after 200th epoch) in Fig. 15 (f).

To further explore the hardware estimation, we extend the simulation to another representative device, EpiRAM [4], whose on-state resistance is relatively smaller (81 kΩ) as compared with 500 kΩ in FeFET. As shown in Fig. 16 (b) and (e), the normalized weight means decrease, while input activities increase before 200th epoch. This different trend of input activities (compare with FeFET-based design) could be caused by the weight initialization in the network, the initial weight means (first 5 epochs) are closer to zero in Fig. 15 (a), but the ones in Fig. 16 (a) are more spread to smaller values.

Due to this, we can find that at first couple of epochs, the weight means decrease (smaller column conductance), but the input activities increase (more activated paths, and thus larger column currents), and overall, the latency shows a slight decreasing trend. The weight means and input activities becomes steady after that, and before 200th epoch, and so as for the latency and dynamic energy. After 200th epoch, both of the weight means and input activates increase, and thus, latency decreases (due to larger column currents), but dynamic energy increases (due to more simultaneously activated transistors in peripheries).

Again, due to the convergence, the energy of weight gradient computation and weight update show similar trends as the FeFET-based design. While the peak energy efficiency shows a more noticeable increase after 200th epoch, this is because the weight update energy contributes more into the total energy. We can find the details from Table I, the write voltage and write pulse width of EpiRAM is much higher than the ones of FeFET, and thus lead to higher weight update energy.

### D. Benchmark Across Technologies

Finally, we benchmark the CIM accelerators for VGG-8 training on CIFAR-10 dataset, with versatile synaptic devices, including the sequential and parallel read-out SRAM-based accelerators at both 7nm and 32nm, and state-of-the-art parallel read-out eNVM-based (including [2][4][6], and [21~23]) accelerators at 32nm.

For versatile analog synaptic devices, the number of conductance varies a lot, i.e. from 32 to 128 levels, thus we change the precision of weights and gradients in the network for different devices according to their device precision (ability to represent high-precision synapse). Thus, overall, we have run different network specs from 5-bit (32 levels) to 7-bit (128 levels). The results show that, with momentum optimization method, and low cycle-to-cycle variation, the FeFET-based based design could achieve quite high accuracy (~91%) even

| VGG-8 on CIFAR10, with Novel Weight Mapping and Dataflow, on-chip training with 256 epochs | | | | | | | | | | |
|---|---|---|---|---|---|---|---|---|---|---|
| Technology node (LSTP) | 7 nm | | 32 nm | | | | | | | |
| Device | SRAM | | SRAM | | Ag:a-Si [21] | PCMO [22] | AlOx/HfO$_2$ [23] | TaOx/HfOx (TsingHua) [2] | EpiRAM [4] | HZO FeFET (NotreDame) [6] |
| # of Conductance States | \ | \ | \ | \ | 97 | 50 | 40 | 128 | 64 | 32 |
| ADC precision | Sequential | 4-bit | Sequential | 4-bit | 6-bit | 5-bit | 5-bit | 6-bit | 6-bit | 5-bit |
| Weight / ΔWeight / Cell Precision | 5-bit / 5-bit / 1-bit | | 5-bit / 5-bit / 1-bit | | 6-bit | 5-bit | 5-bit | 7-bit | 6-bit | 5-bit |
| Ron (Ω) | \ | \ | \ | \ | 26 M | 23 M | 16.9 K | 100 K | 81 K | 500 K |
| On/Off Ratio | \ | \ | \ | \ | 12.5 | 6.84 | 4.43 | 10 | 50.2 | 100 |
| Nonlinearity | \ | \ | \ | \ | 2.4/-4.88 | 3.58/-6.76 | 1.94/-0.61 | 0.04/-0.63 | 0.5/-0.5 | 1.75/1.46 |
| C2C Variation | \ | \ | \ | \ | 3.50% | <1% | 5.00% | 3.70% | 2.00% | <0.5% |
| Write Pulse Voltage | \ | \ | \ | \ | 3.2V/-2.8V | -2V/2V | 0.9V/-1V | 1.6V/1.5V | 5V/-3V | 3.65V/-2.95V |
| Write Pulse Width | \ | \ | \ | \ | 300us/300us | 1ms/1ms | 100us/100us | 50ns/50ns | 5us/5us | 75ns/75ns |
| Area (mm$^2$) | 5.74 | 6.61 | 120.89 | 138.95 | 48.29 | 48.29 | 49.88 | 48.50 | 48.59 | 48.29 |
| Memory Utilization (%) | 94.62% | | | | 88.59% | | | | | |
| Training Accuracy (%) | 91.00% | | | | 49.00% | 56.00% | 37.00% | 81.00% | 85.00% | 91.00% |
| Training Latency (s) / Epoch | 371.17 | 133.45 | 655.32 | 235.75 | 1241.63 | 5795.79 | 611.00 | 177.61 | 193.94 | 176.85 |
| Training Dynamic Energy (J) / Epoch | 110.01 | 93.26 | 149.36 | 95.37 | 92.12 | 92.15 | 93.13 | 92.15 | 92.28 | 92.11 |
| Training Peak Latency (s) / Epoch | 270.70 | 34.95 | 471.43 | 55.33 | 1116.53 | 5670.69 | 485.89 | 52.52 | 68.83 | 51.75 |
| Training Peak Dynamic Energy (J) / Epoch | 9.82 | 2.20 | 113.41 | 15.42 | 9.00 | 9.02 | 10.70 | 9.05 | 9.19 | 8.98 |
| Training Throughput (TOPS) | 0.48 | 1.38 | 0.28 | 0.78 | 0.14 | 0.03 | 0.30 | 1.04 | 0.95 | 1.04 |
| Training Energy Efficiency (TOPS/W) | 1.68 | 1.98 | 1.24 | 1.94 | 2.00 | 2.00 | 1.98 | 2.00 | 2.00 | 2.01 |
| Training Peak Throughput (TOPS) | 0.68 | 5.28 | 0.40 | 3.34 | 0.16 | 0.03 | 0.38 | 3.52 | 2.68 | 3.57 |
| Training Peak Energy Efficiency (TOPS/W) | 18.82 | 83.93 | 1.63 | 11.98 | 20.54 | 20.50 | 17.27 | 20.43 | 20.11 | 20.57 |

Table I. Benchmark results of CIM accelerators training on VGG-8 for CIFAR10, based on SRAM (both sequential and parallel read-out at 7nm and 32nm), and reported "analog" synaptic devices (assumed at 32nm technology). Green bold values shows the good specs and performance.

with 32 levels, thus we set the digital SRAM-based designs to run on 5-bit specs.

From the benchmarking results shown in Table. I, we find that: (1) on-state resistance still plays important role to achieve better hardware performance. Since to avoid large voltage drop, the transistors in 1T1R or peripheral mux have to be sized up for small $R_{on}$, yielding significant area overhead. As a result, it takes longer time to activate the synaptic arrays (due to the increased capacitance loading), adversely increasing latency and lowering throughput. (2) When write pulse width is small, i.e. below a micro second (μs), the weight update will not cause detrimental effects on the speed, as the operation is averaged by batch size. (3) The cycle-to-cycle variation is critical factor for in-situ training accuracy, as large variation could lead to opposite momentum move and prevent the model to learn. A preferred cycle-to-cycle variation is lower than 1%. (4) At same technology node, the SRAM-based designs suffer from leakage and area overhead, while at advanced 7nm, the parallel-read SRAM design still shows superior energy efficiency and throughput.

## V. Conclusion

In this work, we proposed DNN+NeuroSim V2.0, an end-to-end framework to benchmark CIM-based architectures for on-chip training, and support flexible network structures with versatile device options. With introduced behavior model of nonlinearity and asymmetry, device-to-device and cycle-to-cycle variation during weight update, and momentum optimization method to help overcome large asymmetric nonlinearities, it is efficient to investigate the non-ideal properties of analog synaptic devices in in-situ training. From the technological benchmark, it reveals that the desired specs for analog synaptic devices are: low cycle-to-cycle variation (<1%), large on-state resistance (>100 kΩ), small write pulse width (<1μs), with nonlinearity below +3/-3.

## Acknowledgment

This work is supported by ASCENT, one of the SRC/DARPA JUMP centers, NSF/SRC E2CDA program, and NSF-CCF-1903951.